\documentclass[conference]{IEEEtran}

\usepackage{cite}
\usepackage{amsmath,amssymb,amsfonts}
\usepackage{cleveref}
\usepackage{algorithmic}
\usepackage{graphicx}
\usepackage{textcomp}
\usepackage{xcolor}
\usepackage{multirow}
\usepackage{subcaption} 
\usepackage[numbers]{natbib}
\def\BibTeX{{\rm B\kern-.05em{\sc i\kern-.025em b}\kern-.08em
    T\kern-.1667em\lower.7ex\hbox{E}\kern-.125emX}}

\usepackage{todonotes}

\newboolean{showcomments}
\setboolean{showcomments}{true} 
\ifthenelse{\boolean{showcomments}}
{\newcommand{\nb}[2]{
  \fcolorbox{black}{yellow}{\bfseries\sffamily\scriptsize#1}
  {\sf$\blacktriangleright$\textit{#2}$\blacktriangleleft$}
 }
 
}
{\newcommand{\nb}[2]{}
 
}

\newcommand\toolname{NONNO}

\begin{document}


\title{CRATOR: a Dark Web Crawler}

\author{\IEEEauthorblockN{ Jessica De Pascale}
\IEEEauthorblockA{\textit{Tilburg University - JADS}\\
Tilburg, Netherland \\
j.depascale@tilburguniversity.edu}
\and
\IEEEauthorblockN{ Giuseppe Cascavilla}
\IEEEauthorblockA{\textit{TU/e - JADS}\\
Eindhoven, Netherland \\
g.cascavilla@jads.nl}
\and
\IEEEauthorblockN{ Damian~A.~Tamburri}
\IEEEauthorblockA{\textit{TU/e - JADS}\\
Eindhoven, Netherland \\
d.a.tamburri@tue.nl}
\and
\IEEEauthorblockN{ Willem-Jan~Van~Den~Heuvel}
\IEEEauthorblockA{\textit{Tilburg University - JADS}\\
Tilburg, Netherland \\
w.j.a.m.v.d.heuvel@jads.nl}
}

\maketitle

\begin{abstract}

Dark web crawling is a complex process that involves specific methodologies and techniques to navigate the Tor network and extract data from hidden services. This study proposes a general dark web crawler designed to extract pages handling security protocols, such as captchas, efficiently. Our approach uses a combination of seed URL lists, link analysis, and scanning to discover new content. We also incorporate methods for user-agent rotation and proxy usage to maintain anonymity and avoid detection. We evaluate the effectiveness of our crawler using metrics such as coverage, performance and robustness. Our results demonstrate that our crawler effectively extracts pages handling security protocols while maintaining anonymity and avoiding detection. Our proposed dark web crawler can be used for various applications, including threat intelligence, cybersecurity, and online investigations.

\end{abstract}


\section{Introduction}

The dark web is a part of the internet not indexed by search engines and requires special software, such as the Tor browser, to access \cite{chen2012dark}. It is often used for illegal activities, such as selling drugs, weapons, and stolen information.
This part of the web is often associated with illicit activities and is known to host various criminal markets, including illegal drug trading, weapons sales, and stolen data \cite{greenberg2014hacker}. 

Dark web crawling, the act of systematically accessing and collecting data from the dark web, is a complex and challenging task. Unlike the surface web, the dark web lacks structure, making it difficult to discover and navigate its content. Additionally, the anonymity the dark web provides makes it challenging to verify the credibility and authenticity of the information collected.

One common objective of a dark web crawler is to gather information. Because the dark web is largely unregulated and anonymous, it can be a source of valuable data that is not easily accessible through other means. For example, a dark web crawler might be used to search for and collect information such as stolen user credentials, sensitive documents, or other confidential information that has been posted or traded on dark web marketplaces.
Moreover, due to the illegal activities on the dark web, we can collect information to monitor illicit behaviors by looking for trends on specific drugs, armies, or criminal schemes. 
Law enforcement agencies may use dark web crawlers to monitor and gather intelligence on criminal activity such as drug trafficking, human trafficking, and other forms of illicit behavior that take place on the dark web. By collecting data on these activities, law enforcement officials can anticipate and mitigate illicit phenomena or identify and seize the criminal gangs.  
Cybersecurity professionals and companies may also use dark web crawlers to investigate and anticipate potential cyber threats. For example, a dark web crawler could be used to track the activities of hacking groups, monitor the spread of malware or ransomware, or detect data breaches posted on the dark web. This information can then be used to strengthen cybersecurity defenses and protect against future attacks.
An additional application entails examining the user behavior on the dark web or analyzing the materials accessible on it, which can furnish valuable understandings into the subterranean market and the methodologies utilized for illicit activities on the internet.

The implementation of a dark web crawler includes several challenges. The dark web is designed to provide anonymity to users, making it challenging to track and identify the source of information. A dark web crawler must be able to navigate this anonymity and still collect relevant data. Moreover, differently from the surface web, the dark web is unstructured and lacks organization, making it challenging to discover and retrieve specific data.

We aim to address the challenges associated with crawling the dark web and provide researchers and practitioners with a reliable tool for exploring and analyzing dark web content.
Hence, in this study, we propose the development of a general crawler designed to navigate the complex structure of the dark web. Our crawler offers the capability to bypass basic security layers, such as login form pages with simple captchas, and allows for the integration of multiple captchas with real-time manual intervention by the user. The contribution of our work is twofold. Firstly, we propose an architecture specifically tailored for web crawling on the dark web, showcasing the integration of cookie rotation and user-driven manual intervention. Secondly, we have developed a general-purpose crawler for dark web crawling, making it publicly available for use.

To evaluate the performance and effectiveness of our crawler, we conducted an experimentation in which we compared its performance against ACHE \cite{barbosa2007adaptive}, a widely used dark web crawler. The comparison was based on various metrics used to evaluate the crawlers' capabilities. The results clearly indicate that our crawler outperformed ACHE in all the evaluated metrics.

The structure of the paper is organized as follows. 
In Section 2, we discuss the related works on dark web analysis and dark web crawlers, providing insights into the existing research and its limitations. 
Section 3 presents the architecture of our crawler in detail, discussing each layer and its functionalities. 
The evaluation process is described in Section 4, including the metrics used, the dataset employed, and the intended experimentation. 
Section 5 presents the results of our work, highlighting the performance and effectiveness of our crawler compared to other approaches. 
In Section 6, we conclude the paper by summarizing the findings, discussing their implications, and outlining potential avenues for future research.
\section{Related work}

Dark web crawlers are specialized software tools designed to index and collect information from hidden websites and services not accessible through conventional search engines. As the demand for more comprehensive and reliable intelligence on the dark web has increased, several researchers have attempted to develop effective dark web crawlers. In this section, we discuss some of the previous work done in the field of dark web crawling, with a particular focus on the findings of a systematic literature review made by Bergman et al. \cite{bergman2023exploring}.

Bergman et al. conducted a systematic literature review to identify existing dark web crawlers and evaluate their effectiveness. During the review, they identified 34 potential dark web crawlers but found that only four of them 
had publicly available code repositories. The authors conclude the study by highlighting the  absence of open-source
crawlers, and that few researchers mention which open-source crawler was used in their study. Moreover, the experiments showed that the Tor crawler managed to scrape 251 pages in 20 minutes and eight minutes to scrape The Guardian’s clear website of 223 pages with an average of circa 6.4 pages.

Narayanan et al. presents TorBot~\cite{narayanan2020torbot}, an open-source intelligence tool for the dark web that enables users to gather and analyze information from hidden websites and services. TorBot is a Python-based tool that uses the Tor network to crawl the dark web and collect information such as website content, metadata, and links. It has several features, such as keyword-based searching, content analysis, and automated data extraction. The paper discusses the design and implementation of TorBot, including its architecture and components, and provides a detailed evaluation of its performance and capabilities. 

Barbosa et al. introduce ACHE \cite{barbosa2007adaptive}, an adaptive crawler designed to identify entry points to hidden web resources. The hidden web refers to resources that are not accessible through conventional search engines and require specific queries to access. The proposed crawler adapts to the structure of the hidden web by leveraging machine learning techniques to identify entry points and improve its crawling efficiency automatically. The algorithm uses a combination of content-based and link-based analysis to identify hidden web entry points. It continuously updates its search strategy based on the results of previous crawls. The paper describes the design and implementation of the adaptive crawler and evaluates its performance in locating hidden web entry points.

Kalpakis et al. \cite{kalpakis2016interactive} explores an interactive search engine for Home Made Explosives recipes on the Surface and Dark Web. The hybrid architecture combines a dedicated crawler and domain-specific queries, adapting the widely used web crawler, Apache Nutch \cite{khare2004nutch}, to operate on the dark web.

Celestini et al. \cite{celestini2017design} address the need for effective web search and retrieval. It presents a flexible toolkit for structure and content mining on the Web, including the Tor dark Web. The toolkit incorporates web crawling, extraction, indexing, and mining modules. They have customized the well-established web crawler, bUbiNG \cite{boldi2018bubing}, to effectively operate on the dark web. Their adaptation enables the crawler to navigate and retrieve information from the hidden layers of the internet.

In conclusion, we encountered limitations with the availability and suitability of certain crawlers. The TorBot crawler was excluded from our analysis due to outdated external dependencies that rendered it unusable. Additionally, the custom crawlers developed by Kalpakis and Celestini based on Apache Nutch and bUbiNG, respectively, were not publicly released, making their replication challenging. Furthermore, both the crawlers Nutch and bUbiNG were not specifically designed to work with the dark web. Therefore, for our experimentation, we used ACHE (GitHub here \cite{achecrawler}) as a comparison tool due to the fact that, at the moment is the only available and replicable tool suitable for dark web crawling.




\section{Design} \label{sec:design}
\subsection{Architecture}

\begin{figure}[h]
    \centering
    \includegraphics[width=.9\linewidth]{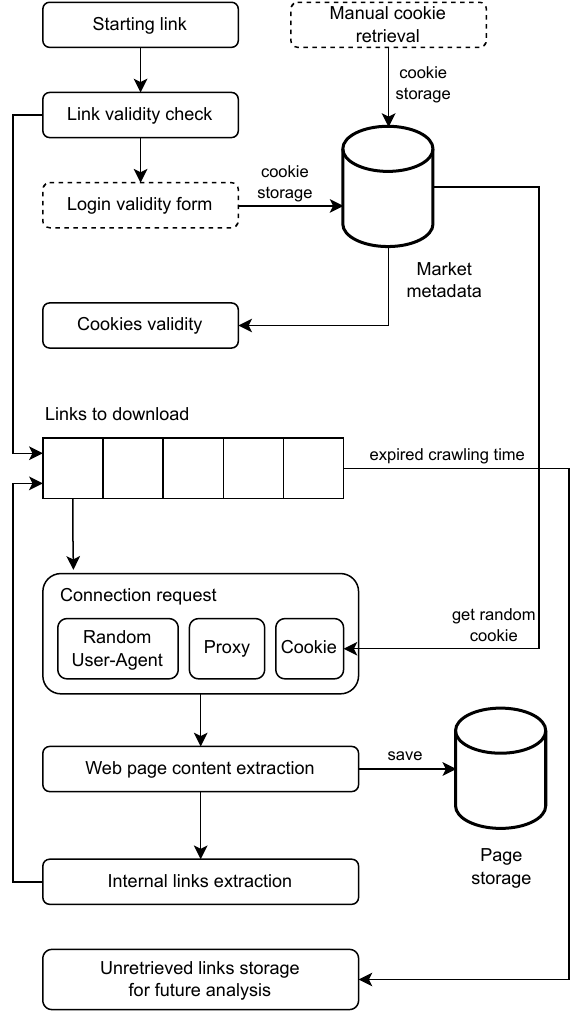}
    \caption{Architecture.}
    \label{fig:crawler-architecture}
\end{figure}

A dark web crawler architecture typically consists of several components that work together to discover hidden web content.
\Cref{fig:crawler-architecture} shows our dark web crawler architecture, giving an overview of the entire crawling process, from the starting link until the content page storage process. The following is a general description of the key components of the architecture. 
Firstly, the crawler receives a list of potential links, namely \textit{starting links}, which it can obtain from public dark web directories or from the user directly. Next, the crawler performs a validity check to ensure that one of these links is accessible. If a valid link is found, it becomes the starting point of the crawling process.

If a website has a security check layer, such as a login page or captcha security check, cookies must be provided to access it after obtaining a valid link. The architecture offers two methods for providing cookies to the crawler. The first method involves human intervention, where the user manually writes one or more cookies for the crawler to use during the crawling process. The second method involves the implementation of a login validity script, which tries to automatically log in and bypass captcha on a dark web link by obtaining login credentials from a Market metadata file. 
In addition, the architecture incorporates a layer for verifying the validity of cookies, in which all collected cookies undergo validation by inspecting for any unexpected redirects. An unexpected redirect is defined as a redirection to a page other than the intended page, such as a response header containing a login link for a homepage link.

After he verification of link and cookie validity, the crawling process starts by downloading the first link along with all internal links, which are links sharing the same domain as the initial link, such as \textit{https://en.wikipedia.org/wiki/IOT} and \textit{https://en.wikipedia.org/wiki/Crawler}, both having the domain \textit{en.wikipedia.org}. For each link, the crawler extract and save the webpage content locally and checks if there are other internal links to put in the download queue. 
The connection module makes use of proxy settings to establish a connection with onion links, cookies to bypass security checks and random user-agent to avoid being identified as a bot.
This process keeps running until reaches a certain exit condition, such as a preset crawling time or a maximum number of links crawled. 

\subsection{Breadth-first Crawling Approach}

There are various strategies for web crawling, and choosing the right strategy is crucial for the success of a crawler.

We decided to adopt a breadth-first crawling strategy because, as proven by Cho et al. \cite{cho1998efficient}, it provides more high-quality pages than other approaches. They use several metrics to define the concept of "quality" of a page, like the number of links that appears during the crawling process or pageRank, a metric used to measure the importance or popularity of web pages based on their incoming links.
In this strategy, the crawler starts with a specific link and then visits all the links on that page before moving to the next page.

\subsection{Link Validity Check}

URL reachability check is an important process that ensures that the crawler can access and retrieve the content of a web page before attempting to extract information from it. This check is necessary because not all URLs that are identified for crawling are necessarily valid or accessible. URLs can become broken due to various reasons such as server errors, domain expiration, or page deletion, making them unreachable and causing errors in the crawling process \cite{liu2011web}.

URL reachability check involves sending a request to the URL or onion link using the HTTP or HTTPS protocol, and checking the response status code returned by the server. A response status code of 200 indicates that the URL is reachable and the server has successfully returned the content of the page. On the other hand, a status code of 404 indicates that the page is not found, while a status code of 503 indicates that the server is temporarily unavailable.

If the crawler determines that a URL is not reachable, it marks the URL as broken or invalid, and remove it from the crawling queue. This helps to prevent unnecessary requests to unreachable URLs, which can slow down the crawling process and waste system resources.

In addition, the crawler may also check if the URL is a duplicate or a mirror of another page, and avoid crawling such pages to prevent duplication of content.


\subsection{Login Forms and Cookies Rotation}

In many cases, web crawlers need to log in to a website to access restricted content. Automating the login process can save time and effort, especially when crawling multiple websites.

To perform an automatic login, a web crawler needs to submit login credentials to the login page of the website. This can be done using HTTP POST requests with the appropriate form data, such as the username and password. The response from the server indicates whether the login was successful or not.

Once the crawler has successfully logged in, it needs to maintain the session to access restricted content. This is usually done using cookies, which are small pieces of data stored on client side and sent with each request to the server. The server uses the cookie to identify the user and maintain the session.

To avoid being detected as a crawler, it is important to rotate the cookies periodically. This can be done by logging out and logging back in again to obtain a new cookie, or by using multiple cookies chosen with a fisher-yates shuffle algorithm. This helps to prevent the website from blocking the crawler or treating it as suspicious activity.

It's worth noting that some websites may have additional security measures in place, such as CAPTCHAs or two-factor authentication. These can make the automatic login process more challenging and may require additional coding or human intervention.

Our architecture offers the flexibility of managing cookies manually or automatically. This is achieved through an automated login script that logs into the starting link pages that require username and password or by including a list of cookies in the market metadata file.

\subsection{Captcha detection}

Captcha detection is an important aspect of web crawling, as many websites use captchas to prevent automated bots from accessing their content. Captchas are usually presented as a challenge-response test, where the user is asked to perform a task that is easy for humans but difficult for machines, such as identifying distorted text, selecting images that match a certain criteria, or solving a mathematical problem.

Detecting captchas in a web crawler can be challenging, as captchas can appear in different forms and locations on the page, and can be triggered by various factors, such as the number of requests, the frequency of requests, or the IP address of the crawler. To detect captchas, web crawlers need to analyze the page structure and content, and look for patterns that indicate the presence of a captcha.

\subsection{Connection Setup}

On the surface web, connection requests are typically made using the standard HTTP or HTTPS protocols. This involves sending a request to a web server and receiving a response that includes the requested content. The server may also set cookies or other tracking data that the client can store and use in future requests.

In contrast, the dark web operates on a different protocol called Onion Routing \cite{goldschlag2005hiding}, which uses several layers of encryption to anonymize communication between clients and servers. To make connection requests on the dark web, the crawler must be configured to use the Tor network, which provides a layer of encryption and routing that obscures the origin and destination of requests, and a proxy server. A proxy server acts as an intermediary between the client and the server, forwarding requests and responses and providing an additional layer of anonymity. The proxy server can be either a public or a private one.

A further technique useful to help avoid being detected and blocked by servers is the user agent rotation. A user agent is a string that identifies the type of browser and operating system being used to make the request. By rotating the user agent, a crawler can simulate human-like behavior and avoid being detected as an automated bot.

One of the main differences between surface web and dark web requests is the domain resolution process. On the surface web, domain names are translated into IP addresses using a centralized DNS resolution system. However, the dark web uses a decentralized naming system called Tor Hidden Services, which allows websites to have .onion domain names that are not registered with standard DNS servers. Python scripts that are designed to make connection requests on the dark web must be configured to use Tor Hidden Services to resolve domain names and connect to the appropriate servers.

\subsection{Stop Criteria}

A crawler uses exit conditions to determine when the crawling process should stop. There are several exit conditions that can be used depending on the goals of the crawler and the resources available. Our algorithm employs the following stop criteria:

\begin{itemize}
    \item[-] \textbf{Max depth}: the crawler stops if it reaches a certain level of depth. This criteria is useful when you want to limit the amount of resources used during the crawling process.
    \item[-] \textbf{Max number of link downloaded}: one common exit condition is to stop crawling after a certain number of links have been visited. This approach is useful when you want to limit the amount of data collected or when you are crawling a large website and want to ensure that the process does not continue indefinitely.
    \item[-] \textbf{Time limit}: it is important to stop crawling after a certain amount of time has passed. Considering the dynamic nature of the dark web, where lots of website are frequently updated, having a stop criteria condition ensures that the data collected is up-to-date.
    \item[-] \textbf{Data crawled}: the last stop criteria is determined by whether or not the crawler has collected all the targeted links.
\end{itemize}

\section{Evaluation}

In this section, we describe the evaluation criteria and the case scenaario used to evaluate a dark web crawler.

\subsection{Case Scenario}

For the evaluation of our dark web crawling methodology, we chose to analyze a prominent dark web marketplaces, \textit{Cocorico Market}. This marketplaces is known for hosting a wide range of illicit activities, including the sale of drugs, weapons, counterfeit items, stolen data, and various other illegal goods and services.

\subsection{Metrics} \label{subsec: metrics}

\subsubsection{Coverage}
The coverage of a crawler refers to the extent to which it is able to successfully and comprehensively extract all of the desired data from a given set of sources. Measuring the coverage is important for evaluating the effectiveness of the crawler and ensuring that it is meeting its intended goals.
One way to measure the coverage of a crawler is by calculating the ratio of the number of downloaded pages to the total number of web pages of a website \cite{girardi2006web}. Since the total number of pages of a website is not known a priory, an approach to estimating the coverage of a crawler is to compare the number of pages downloaded by the crawler to the total number of pages downloaded by other crawlers.

\subsubsection{Performance}
Performance is a critical factor in evaluating the effectiveness of a web crawler, and there are several key metrics that can be used to measure its performance. Two common metrics used to evaluate the performance of a crawler are the number of pages downloaded per minute and the execution time of each crawler \cite{bharati2013higwget}.

The number of pages downloaded per minute is a measure of how quickly a crawler can traverse the web and retrieve content. This metric is essential for ensuring that the crawler can keep up with the constant stream of new content being added to the web. A high rate of page downloads per minute can be an indication of a well-performing crawler, while a low rate may suggest that the crawler is struggling to keep up with the volume of content.

In evaluating the performance of a dark web crawler, the execution time is a fundamental metric due to its direct correlation with the efficiency and responsiveness of the crawler in retrieving web content. A shorter execution time indicates a more agile and responsive crawler. As web crawling tasks often involve processing vast amounts of data distributed across numerous web pages, minimizing execution time is important to ensure timely data retrieval and processing.

\subsubsection{Robustness}
In the context of web crawling, robustness refers to the ability of a crawler to continue operating effectively and efficiently even when faced with unexpected or challenging conditions. A robust crawler is able to handle a variety of situations, including network connectivity issues, site changes, and other potential obstacles, without experiencing significant disruptions or failures. Robustness is an important characteristic of a successful crawler, as it ensures that the crawler is able to continue operating effectively over time, even in the face of changing conditions and challenges.
The robustness can be measured by counting the number of error happen during the crawling process \cite{girardi2006web}.

\subsection{Experimentation}

To evaluate the effectiveness of our dark web crawler, we conduct experiments comparing our crawler with the dark web crawler ACHE. We select this crawler based on its prominence in the field and their ability to navigate the dark web.

The goal of the experimentation is to assess the performance of our crawler in terms of coverage, performance, and robustness. We aim to determine how well our crawler gather data from the dark web, enlightning the strengths and weaknesses of our methodology in capturing data from Cocorico Market. The evaluation metrics provided a comprehensive framework for assessing the performance of the crawlers and determining the suitability of our approach in the context of dark web crawling.

The experimental setup involved running each crawler separately on identical hardware infrastructure, using the same configurations and parameters. To ensure a fair comparison among the crawlers, we established a specific time frame of 10 hours for the crawling process. The experimentation was repeated 10 times for each level of depth, ranging from depth 1 to depth 4. Subsequently, we calculated the mean and standard deviation of each metric involved in the analysis.
\section{Results}

In this section, we present the results of the experimentation conducted on both dark web crawlers \textit{ACHE} and \textit{CRATOR}.

\subsection{Coverage}

\begin{table}[!h]
\centering
\renewcommand{\arraystretch}{1.5}
\resizebox{\linewidth}{!}{
    \begin{tabular}{llll|lll}
        \hline
        & \multicolumn{3}{c}{\textbf{ACHE}} & \multicolumn{3}{c}{\textbf{CRATOR}} \\
        \hline
        & identified & success & rate & identified & success & rate \\ 
        \hline
        \multicolumn{7}{c}{\textbf{DEPTH 1}}\\
        \hline
        mean & 117.0 & 110.5 & 0.944 & 115.9 & 115.9 & 1 \\ \hline
        std & 0.0 & 19.506 & 0.167 & 0.316 & 0.316 & 0 \\ \hline
        \multicolumn{7}{c}{\textbf{DEPTH 2}}\\
        \hline
        mean & 539.3 & 516.1 & 0.829 & 622.2 & 622.2 & 1 \\ \hline
        std & 214.925 & 246.84 & 0.36 & 12.2 & 12.2 & 0 \\ \hline
        \multicolumn{7}{c}{\textbf{DEPTH 3}}\\
        \hline
        mean & 1757.2 & 1395 & 0.763 & 2266.5 & 2266.5 & 1 \\ \hline
        std & 619.048 & 616.559 & 0.154 & 79.271 & 79.271 & 0 \\ \hline
        \multicolumn{7}{c}{\textbf{DEPTH 4}}\\
        \hline
        mean & 2784.1 & 2210.5 & 0.705 & 4125.7 & 4125.7 & 1 \\ \hline
        std & 1148.065 & 1110.328 & 0.287 & 193.82 & 193.82 & 0 \\ \hline
        \hline
    \end{tabular}
}
\vspace{.1cm}
\caption{Results of the coverage, as number of success links downloaded by each approach, and relative coverage (rate) of CRATOR and ACHE by analysing Cocorico market with 10 iteration and 4 level of depth.}
\label{tab:res-coverage}
\end{table}

\begin{figure*}[htbp]
    \centering
    \begin{subfigure}[b]{0.49\textwidth}
        \centering
        \includegraphics[width=\textwidth]{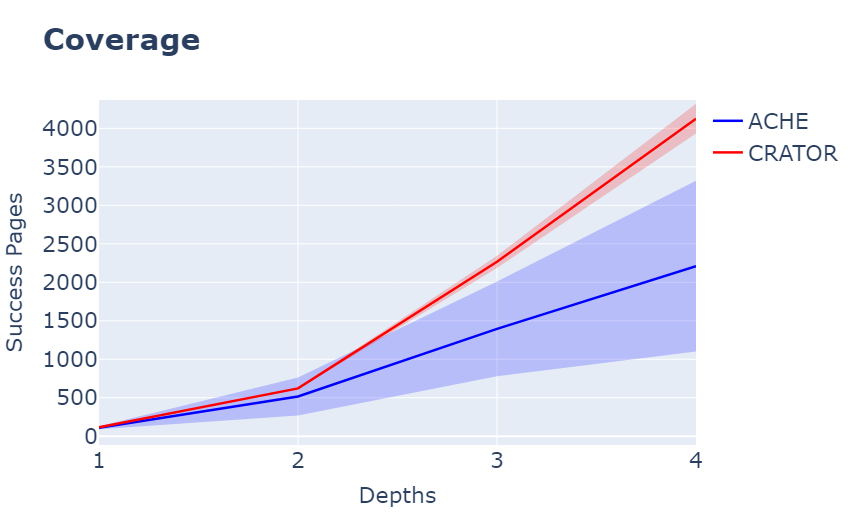}
        \caption{Caption for Figure 1}
        \label{fig:res-absolute-coverage}
    \end{subfigure}
    \hfill
    \begin{subfigure}[b]{0.49\textwidth}
        \centering
        \includegraphics[width=\textwidth]{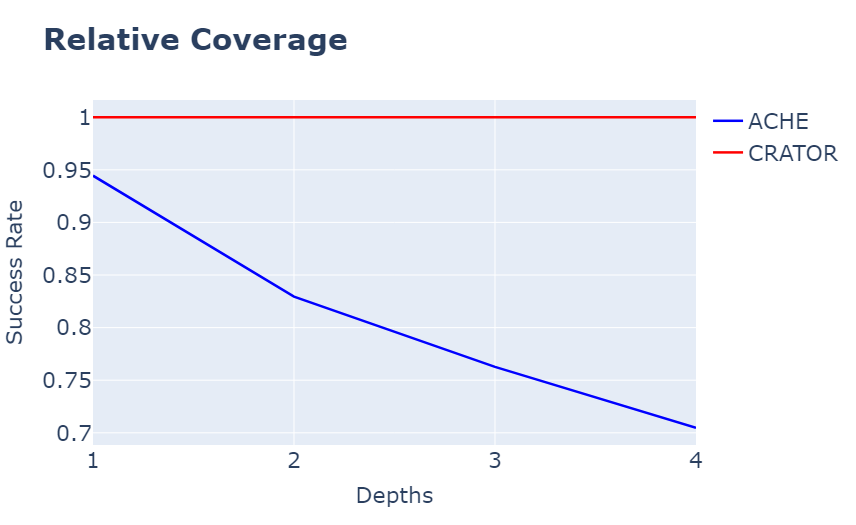}
        \caption{Caption for Figure 2}
        \label{fig:res-relative-coverage}
    \end{subfigure}
    \caption{Caption for both figures}
    \label{fig:res-coverage}
\end{figure*}

In the analysis of the dark web marketplace \textit{Cocorico Market} to validate the coverage, we considered two coverage metrics as explained in Section \ref{subsec: metrics}. Coverage is defined as the number of pages downloaded, while relative coverage is the success rate of a crawler calculated as pages downloaded divided by pages found.

Our tool, \textit{CRATOR}, demonstrated remarkable effectiveness in downloading all the pages it found. On the other hand, \textit{ACHE} achieved a maximum coverage of 0.944 with a level of depth 1 (see results in \cref{tab:res-coverage}). When considering relative coverage, it seems that \textit{CRATOR} outperforms \textit{ACHE} consistently.

This metric provides insights into the internal coverage of the tool, as it is based on the links discovered during crawling. Hence, it lacks knowledge of the real size of the marketplace. By utilizing coverage metrics, we gain a better understanding of the capabilities of both tools in downloading all the pages within the marketplace.

From the results presented in \cref{tab:res-coverage} and the visual representation in \cref{fig:res-absolute-coverage}, it is evident that \textit{CRATOR} consistently outperforms \textit{ACHE} across all scenarios (levels of depth 1, 2, 3, 4). Furthermore, the standard deviation outlined in \cref{fig:res-absolute-coverage} indicates that \textit{CRATOR} exhibits higher reliability across different iterations. The lower standard deviation of \textit{CRATOR} suggests that the results achieved with different executions are more consistent compared to \textit{ACHE}.

\subsection{Performance}

\begin{table}[h]
\centering
\renewcommand{\arraystretch}{1.5}
\resizebox{\linewidth}{!}{
    \begin{tabular}{@{}llllll@{}}
    \hline
     & & \textbf{DEPTH 1} & \textbf{DEPTH 2} & \textbf{DEPTH 3} & \textbf{DEPTH 4} \\ \hline
    \multicolumn{1}{c}{\multirow{2}{*}{\textbf{ACHE}}} & mean & 268.373 & 1114.383 & 3612.991 & 5615.148 \\
    \multicolumn{1}{c}{} & std & 85.021 & 416.752 & 1220.657 & 2311.664 \\
    \multirow{2}{*}{\textbf{CRATOR}} & mean & 173.3 & 939.2 & 3420.6 & 6356.2 \\
     & std & 1.418 & 18.103 & 119.295 & 350.985 \\
    \hline
    \end{tabular}
}
\vspace{.1cm}
\caption{Execution time (seconds)}
\label{tab:results-exec_time}
\end{table}

\begin{table}[!h]
\centering
\renewcommand{\arraystretch}{1.5}
\resizebox{\linewidth}{!}{
    \begin{tabular}{@{}llllll@{}}
    \hline
     & & \textbf{DEPTH 1} & \textbf{DEPTH 2} & \textbf{DEPTH 3} & \textbf{DEPTH 4} \\ \hline
    \multicolumn{1}{c}{\multirow{2}{*}{\textbf{ACHE}}} & mean & 22.815 & 24.767 & 28.322 & 26.602 \\
    \multicolumn{1}{c}{} & std & 4.317 & 8.666 & 1.742 & 8.998 \\
    \multirow{2}{*}{\textbf{CRATOR}} & mean & 28.975 &37.276 & 39.076 & 38.623 \\
     & std & 0.079 & 0.929 & 0.280 & 2.015 \\
    \hline
    \end{tabular}
}
\vspace{.1cm}
\caption{Pages per minute.}
\label{tab:results-ppm}
\end{table}

\begin{figure*}[htbp]
    \centering
    \begin{subfigure}[b]{0.49\textwidth}
        \centering
        \includegraphics[width=\textwidth]{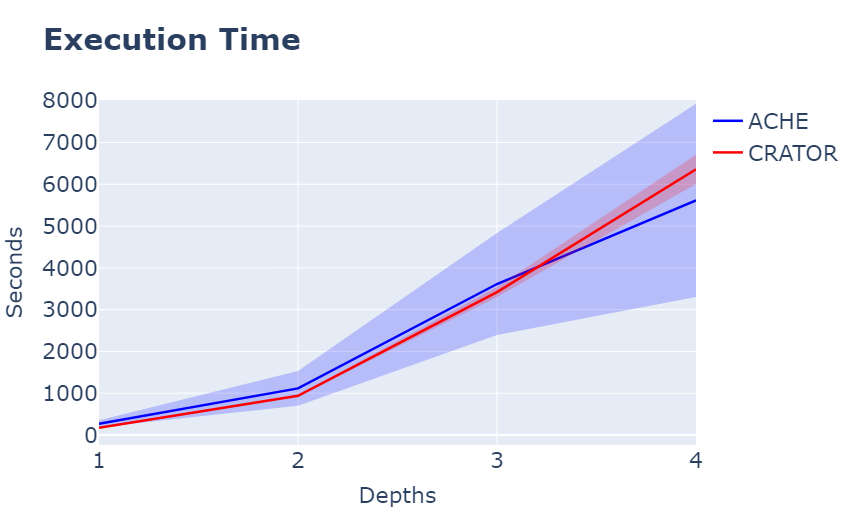}
        \caption{Execution time.}
        \label{fig:res-execution-time}
    \end{subfigure}
    \hfill
    \begin{subfigure}[b]{0.49\textwidth}
        \centering
        \includegraphics[width=\textwidth]{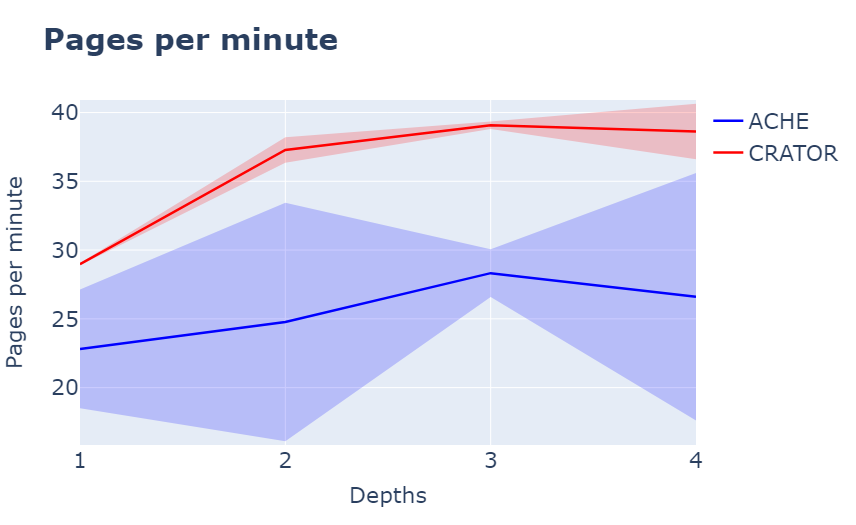}
        \caption{Pages per minute.}
        \label{fig:res-ppm}
    \end{subfigure}
    \caption{Performance metrics.}
    \label{fig:res-performance}
\end{figure*}

As outlined in \cref{subsec: metrics}, we evaluate our tool regarding the performance metric using two key metrics: execution time and the number of pages downloaded per minute. \Cref{fig:res-execution-time} illustrates the execution time of both approaches. It is evident from the figure that \textit{CRATOR} exhibits lower execution times compared to \textit{ACHE} for the first three levels of depth. However, \textit{ACHE} outperforms \textit{CRATOR} with the last level of depth.

To understand this discrepancy, we analyse the second performance metric. \Cref{fig:res-ppm} presents the pages downloaded per minute by each tool for each level of depth. It is evident that \textit{CRATOR} consistently outperforms \textit{ACHE} in every scenario. Thus, despite the higher execution time of \textit{CRATOR} at level of depth 4, it is clear that \textit{CRATOR} downloaded more pages, as confirmed by the results presented in \cref{tab:res-coverage}.

\subsection{Robustness}

\begin{table}[!h]
\centering
\renewcommand{\arraystretch}{1.5}
\resizebox{\linewidth}{!}{
    \begin{tabular}{llll|lll}
        \hline
        & \multicolumn{3}{c}{\textbf{ACHE}} & \multicolumn{3}{c}{\textbf{CRATOR}} \\
        \hline
        & identified & failed & rate & identified & failed & rate \\ 
        \hline
        \multicolumn{7}{c}{\textbf{DEPTH 1}}\\
        \hline
        mean & 117.0 & 6.5 & 0.056 & 115.9 & 0 & 0 \\ \hline
        std & 0.0 & 19.506 & 0.167 & 0 & 0 & 0 \\ \hline
        \multicolumn{7}{c}{\textbf{DEPTH 2}}\\
        \hline
        mean & 539.3 & 23.2 & 0.171 & 622.2 & 0 & 0 \\ \hline
        std & 214.925 & 66.344 & 0.36 & 12.2 & 0 & 0 \\ \hline
        \multicolumn{7}{c}{\textbf{DEPTH 3}}\\
        \hline
        mean & 1757.2 & 362.2 & 0.237 & 2266.5 & 0 & 0 \\ \hline
        std & 619.048 & 219.622 & 0.154 & 79.271 & 0 & 0 \\ \hline
        \multicolumn{7}{c}{\textbf{DEPTH 4}}\\
        \hline
        mean & 2784.1 & 573.6 & 0.295 & 4125.7 & 0 & 0 \\ \hline
        std & 1148.065 & 475.628 & 0.287 & 193.82 & 0 & 0 \\ \hline
        \hline
    \end{tabular}
}
\vspace{.1cm}
\caption{Failure rate of \toolname and ACHE by analysing Cocorico market with 10 iteration and 4 level of depth.}
\label{tab:results-failure-rate}
\end{table}

\begin{figure*}[!h]
    \centering
    \includegraphics[width=.8\linewidth]{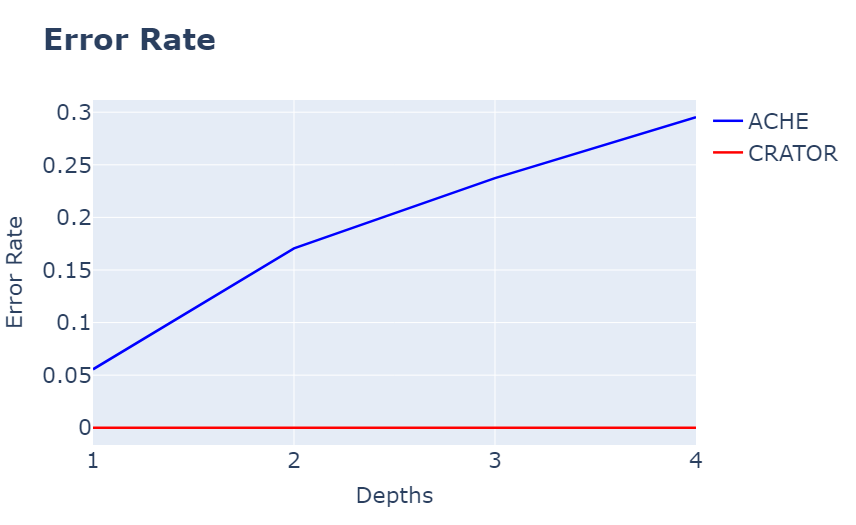}
    \caption{Error rate.}
    \label{fig:res-error}
\end{figure*}

As discussed in \cref{subsec: metrics}, we evaluate robustness using the failure rate metric, which is expressed as the number of pages not downloaded over the total number of pages found. \Cref{tab:results-failure-rate} presents the number of pages failed, with the failure rate expressed as a value between 0 and 1.

In \cref{fig:res-error}, we observe that our tool, \textit{CRATOR}, successfully downloaded all the pages found, demonstrating a failure rate of 0. This is in contrast to \textit{ACHE}, which exhibits a failure rate ranging between 0.05 and 0.29. These results underscore the effectiveness of the techniques employed to conceal our crawler's activities on the dark web, as outlined in \cref{sec:design}. Specifically, techniques such as proxy usage, IP rotation, and cookie rotation have proven to be efficient in crawling the dark web while minimizing the risk of detection and failure.

The results obtained from the evaluation highlight the strengths of \textit{CRATOR} in terms of coverage, performance, and robustness. It exhibited the ability to download a higher number of pages, achieved a higher average of pages per minute, and demonstrated lower error rates.
\section{Conclusion and Future Work}

In this study, we presented \textit{CRATOR}, a dark web crawler that offers an efficient and robust approach for capturing data from dark web marketplaces. We outlined the architecture of \textit{CRATOR}, emphasizing its security features, including an automatic login form, cookies rotation, and random user agent implementation, which enhance its ability to handle security checks prevalent in the dark web.

To validate the effectiveness of our crawler, we conducted an experimentation comparing \textit{CRATOR}~with an established dark web crawler, \textit{ACHE}. The comparison was based on metrics such as coverage, performance, and robustness. The results demonstrated that \textit{CRATOR} outperformed \textit{ACHE} in terms of coverage, downloading a greater number of pages within the specified timeframe. It also exhibited a higher page per minute rate, indicating its efficiency in navigating and retrieving information. Additionally, \textit{CRATOR} raised fewer errors, highlighting its robustness and ability to overcome common obstacles encountered in the dark web.

As a future work, we plan to conduct a more exhaustive experimentation, focusing on daily analysis of dumps and incorporating different dark web marketplaces with varying security layers. This expanded evaluation will allow us to assess the usability and performance of \textit{CRATOR} in a broader range of scenarios, further validating its capabilities as a reliable dark web crawler.

Our study demonstrates the potential of \textit{CRATOR} as an effective tool for dark web crawling, offering improved coverage, performance, and robustness compared to existing approaches. With further research and refinement, \textit{CRATOR} has the potential to contribute significantly to the exploration and understanding of the dark web ecosystem.


\bibliographystyle{IEEEtranN}
\bibliography{references.bib}

@book{chen2012dark,
  title={Dark web exploring and data mining the dark side of the web},
  author={Chen, Hsinchun},
  year={2012},
  publisher={Springer}
}

@article{greenberg2014hacker,
  title={Hacker lexicon: what is the dark web?},
  author={Greenberg, Andy},
  journal={Wired. http://www. wired. com/2014/11/hacker-lexicon-whats-dark-web [dostkep 6.02. 2017]},
  year={2014}
}

@article{cho1998efficient,
  title={Efficient crawling through URL ordering},
  author={Cho, Junghoo and Garcia-Molina, Hector and Page, Lawrence},
  journal={Computer networks and ISDN systems},
  volume={30},
  number={1-7},
  pages={161--172},
  year={1998},
  publisher={Elsevier}
}

@article{liu2011web,
  title={Web crawling},
  author={Liu, Bing and Liu, Bing and Menczer, Filippo},
  journal={Web Data Mining: Exploring Hyperlinks, Contents, and Usage Data},
  pages={311--362},
  year={2011},
  publisher={Springer}
}

@inproceedings{goldschlag2005hiding,
  title={Hiding routing information},
  author={Goldschlag, David M and Reed, Michael G and Syverson, Paul F},
  booktitle={Information Hiding: First International Workshop Cambridge, UK, May 30--June 1, 1996 Proceedings},
  pages={137--150},
  year={2005},
  organization={Springer}
}

@article{bergman2023exploring,
  title={Exploring Dark Web Crawlers: A systematic literature review of dark web crawlers and their implementation},
  author={Bergman, Jesper and Popov, Oliver B},
  journal={IEEE Access},
  year={2023},
  publisher={IEEE}
}

@article{girardi2006web,
  title={Web crawlers compared},
  author={Girardi, Christian and Ricca, Filippo and Tonella, Paolo},
  journal={International Journal of Web Information Systems},
  year={2006},
  publisher={Emerald Group Publishing Limited}
}

@article{bharati2013higwget,
  title={HIGWGET-A Model for Crawling Secure Hidden WebPages},
  author={Bharati, KF and Premchand, P and Govardhan, A},
  journal={International Journal of Data Mining \& Knowledge Management Process},
  volume={3},
  number={2},
  pages={23},
  year={2013},
  publisher={Academy \& Industry Research Collaboration Center (AIRCC)}
}

@inproceedings{narayanan2020torbot,
  title={Torbot: open source intelligence tool for dark web},
  author={Narayanan, PS and Ani, R and King, Akeem TL},
  booktitle={Inventive Communication and Computational Technologies: Proceedings of ICICCT 2019},
  pages={187--195},
  year={2020},
  organization={Springer}
}

@inproceedings{barbosa2007adaptive,
  title={An adaptive crawler for locating hidden-web entry points},
  author={Barbosa, Luciano and Freire, Juliana},
  booktitle={Proceedings of the 16th international conference on World Wide Web},
  pages={441--450},
  year={2007}
}

@article{boldi2018bubing,
  title={BUbiNG: Massive crawling for the masses},
  author={Boldi, Paolo and Marino, Andrea and Santini, Massimo and Vigna, Sebastiano},
  journal={ACM Transactions on the Web (TWEB)},
  volume={12},
  number={2},
  pages={1--26},
  year={2018},
  publisher={ACM New York, NY, USA}
}

@inproceedings{kalpakis2016interactive,
  title={Interactive discovery and retrieval of web resources containing home made explosive recipes},
  author={Kalpakis, George and Tsikrika, Theodora and Iliou, Christos and Mironidis, Thodoris and Vrochidis, Stefanos and Middleton, Jonathan and Williamson, Una and Kompatsiaris, Ioannis},
  booktitle={Human Aspects of Information Security, Privacy, and Trust: 4th International Conference, HAS 2016, Held as Part of HCI International 2016, Toronto, ON, Canada, July 17-22, 2016, Proceedings 4},
  pages={221--233},
  year={2016},
  organization={Springer}
}

@article{khare2004nutch,
  title={Nutch: A flexible and scalable open-source web search engine},
  author={Khare, Rohit and Cutting, Doug and Sitaker, Kragen and Rifkin, Adam},
  journal={Oregon State University},
  volume={1},
  pages={32--32},
  year={2004},
  publisher={Citeseer}
}

@inproceedings{celestini2017design,
  title={Design, implementation and test of a flexible tor-oriented web mining toolkit},
  author={Celestini, Alessandro and Guarino, Stefano},
  booktitle={Proceedings of the 7th International Conference on Web Intelligence, Mining and Semantics},
  pages={1--10},
  year={2017}
}

@misc{achecrawler, 
title={ACHE Focused Crawler}, 
author={{Aécio Santos and Kien Pham}},
year={2023}, 
howpublished = {https://github.com/VIDA-NYU/ache}
}

\end{document}